\documentclass[12pt,a4paper]{article}
\usepackage{a4wide}
\usepackage{amsfonts}
\usepackage{amssymb}
\usepackage{amsmath}
\usepackage{cancel}
\usepackage{latexsym}
\usepackage{epsf}
\usepackage{amssymb}
\usepackage{cancel}
\usepackage{slashed}
\begin{document}
\def\be{\begin{equation}}
\def\ee{\end{equation}}
\def\bea{\begin{eqnarray}}
\def\eea{\end{eqnarray}}

\newcommand{\viel}[4]{e_{\phantom{#1}#2}^{\phantom{#3}#4}}

\def\pd{\partial}
\def\a{\alpha}
\def\b{\beta}
\def\g{\gamma}
\def\d{\delta}
\def\m{\mu}
\def\n{\nu}
\newcommand{\fsl}{{\hspace{-7pt}\slash}}
\newcommand{\gsl}{{\hspace{-5pt}\slash}}
\newcommand{\dslash}{\pd\fsl}
\newcommand{\pslash}{p\gsl }
\newcommand{\Aslash}{A\gsl }
\newcommand{\qslash}{q\gsl }
\newcommand{\Lslash}{\Lambda\gsl }
\newcommand{\kuslash}{k_1\gsl }
\newcommand{\kdslash}{k_2\gsl }
\newcommand{\Dslash}{D\fsl}
\def \h{\mathcal{H}}
\def \hh{\mathcal{G}}
\def\t{\tau}
\def\p{\pi}
\def\th{\theta}
\def\l{\lambda}
\def\O{\Omega}
\def\r{\rho}
\def\s{\sigma}
\def\e{\epsilon}
  \def\scri{\mathcal{J}}
\def\cM{\mathcal{M}}
\def\tcM{\tilde{\mathcal{M}}}
\def\RR{\mathbb{R}}

\hyphenation{re-pa-ra-me-tri-za-tion}
\hyphenation{trans-for-ma-tions}
\hyphenation{a-cce-le-ra-tion}


\begin{flushright}
IFT-UAM/CSIC-08-15\\
hep-th/yymmnnn\\
\end{flushright}

\vspace{1cm}

\begin{center}

{\bf\Large  Very Special (de Sitter) Relativity}

\vspace{.5cm}

{\bf Enrique \'Alvarez and Roberto Vidal}

\vspace{.3cm}

\vskip 0.4cm  
 
{\it  Instituto de F\'{\i}sica Te\'orica UAM/CSIC, C-XVI,
and  Departamento de F\'{\i}sica Te\'orica, C-XI,\\
  Universidad Aut\'onoma de Madrid 
  E-28049-Madrid, Spain }

\vskip 0.2cm

\vskip 1cm

\end{center}
{\bf Abstract}
The effects of a non-vanishing value for the cosmological constant in the scenario of Lorentz symmetry breaking recently proposed by Cohen and Glashow (which they denote as Very Special Relativity) are explored and observable consequences are pointed out.


\begin{quote}

\end{quote}


\newpage

\setcounter{page}{1}
\setcounter{footnote}{1}
\tableofcontents
\newpage

\section{Introduction}
There is a growing consensus that the most likely interpretation of the observed acceleration of the Universe (cf. \cite{Komatsu} for a recent reference) is the presence of a cosmological constant, of mass dimension two, whose energy scale corresponds roughly to the inverse of the Hubble radius, $R_H\sim 10^{10}\textrm{ ly}$. This means $\displaystyle\l\sim \frac{1}{R_H^2}\sim 10^{-65}\textrm{ eV}^2$, so that the corresponding mass scale is the {\em Hubble mass},
 $M_\l\equiv \l^{1/2}\sim 10^{-33}\textrm{ eV}$.\\

When interpreted as a vacuum energy density, the quantity of interest is $M_p^2 M^2_\l\equiv M_{DE}^4$, (where $M_P$ is the Planck mass). This yields the {\em dark energy} scale, $M_{DE}\sim 10^{-3}\textrm{ eV}$.\\

It should be remarked that these two scales, $M_\l$ and $M_{DE}$ are quite different; 
the former is the one that determines the corresponding de Sitter geometry, whereas the latter is related to possible dynamical effects of the matter.\\

The propagation of all massive particles is then affected by powers of $\xi$, 
the dimensionless ratio of the mass scale associated to the particle and the Hubble scale:
\be
\xi\equiv\frac{M_\l}{m}
\ee
Massless particles are also affected in subtle ways, and the only known particle for which
a large value of $\xi$ is not excluded is the lightest neutrino.\\

A new line of thought concerning the neutrino masses has been recently put forward by Cohen 
and Glashow \cite{Cohen}. They disposed of the hypothesis that the full rotation group 
ought to  be preserved in any realistic scenario of Lorentz symmetry violation. In that 
way they pointed out the possibility of  a breaking from $SO(1,3)$ to a proper subgroup, 
a four-parameter subgroup. The scenario in which the fundamental symmetry of Nature is 
just $SIM(2)$ plus translations was dubbed by them Very Special Relativity (VSR), and 
in this framework it is possible to postulate a new origin for the neutrino masses.\\

A fascinating point of this approach is that Lorentz symmetry breaking appears inextricably 
entangled with parity breaking, in the sense that the addition of just parity to VSR is 
enough to restore the full Lorentz symmetry. Another curious consequence is that there are 
no new invariant tensors for $SIM(2)$, so that any local term that can be written 
in the Lagrangian enjoys enhanced full Lorentz symmetry. It seems that intrinsic Lorentz
 breaking effects in this scheme must be both parity violating and non local in nature.\\

In this paper, the  generalization of VSR ideas to de Sitter spacetime is studied. This can be done in two different ways, depending on whether preeminence is given to boost invariance or else to rotation invariance. The nonlocal  equation of motion postulated by Cohen and Glashow can easily be derived from a local lagrangian, with auxiliary fields. We generalize it to the de Sitter space,
and we study its symmetries in differnt cases with some care.

\section{Breaking de Sitter invariance}

The difference between the Poincare and the de Sitter groups are the non-vanishing commutation relations between translations in the latter. Then, one possible approach in order to break the de Sitter invariance is to start with the $SIM(2)$ algebra\footnote{Remember that: $T_1=K_1-J_2$, $T_2=K_2+J_1$, and $A,B=1,2$. Along the full paper only non vanishing commutators will appear.}:
\begin{align}
 & [J_3,T_A]=i\epsilon_{AB}T_B\nonumber\\
 & [K_3,T_A]=iT_A
\end{align}
and try to include as many translations as we can. Let us coin the provisional name $\Lambda$VSR for the resulting group.\\

Given the fact that:
\begin{align}
&[K_3,H]=iP_3\textrm{ , }[K_3,P_3]=i H \nonumber\\
& \left[T_A,P_3\right] =-iP_A\textrm{ , }[T_A,P_B]=i P_+ \delta_{AB}\textrm{ , }[T_A,H]=iP_A\nonumber\\
&\left[J_3,P_A\right]=i\epsilon_{AB}P_B\nonumber\\
&\left[P_i,P_j\right]=i\e_{ijk}J_k\textrm{ , }[H,P_i]=iK_i\textrm{ ,}
\end{align}
where $P_\pm\equiv H\pm P_3$, it easily follows that we can include only $P_+$ 
as a (hamiltonian) traslation,
\begin{equation}
[T_A,P_+]=0\textrm{ , }[K_3,P_+]=iP_+\textrm{ , }[J_3,P_+]=0
\end{equation}
If we include $P_A$, then we have to include $K_A$ as well, because $[H,P_A]=K_A$. 
If we include $P_3$, this forces to include $P_A$,  owing to $[T_A,P_3]=iP_A$. Even if we include $H$ by itself, we are forced to include $P_A$ again, because $[T_A,H]=iP_A$.\\

The appropiate subgroup of de Sitter that corrresponds to $\Lambda$VSR is then generated by $\left\{T_A,K_3,J_3,P_+\right\}$:
 \bea
&&[K_3,T_A]=i T_A\nonumber\\
&&[J_3,T_A]=i\e_{AB}T_B\nonumber\\
&&[K_3,P_+]=iP_+
\eea
This is obviously a subgroup of the Lorentz group as well, the only thing that is specific to de Sitter in this context is the inability to consider all components of the momentum as quantum numbers.\\

To summarize, the net effect of the comological constant, is to reduce the number of available generators from eight ($SIM(2)$ plus
translations) down to five. That is, if we follow the same philosophy of breaking the rotations and keeping the boost $K_3$ that lead to VSR in flat space. But there is another possibility. Since $SIM(2)$ is defined by the requirement of leaving invariant a certain null direction, and since the de Sitter group is the five dimensional equivalent of Lorentz group, we can exploit this analogy.\\

Let us split the generators into four {\em boosts} $M_{0I}\equiv K_I$ and six {\em rotations} $M_{IJ}$
where the five-dimensional spatial indices run from $I,J,\ldots=1,2,3,4$. Next define, for the four-dimensional spatial indices $i,j,\ldots=1,2,3$, $M_{ij}=\e_{ijk} L_k$ and $M_{4i}=N_i$. The commutators read:
\begin{align}
&[K_4,K_i]=-i N_i\textrm{ , }[K_4,N_i]=-iK_i\textrm{ , }\nonumber\\
&[K_i,K_j]=-i \e_{ijk}L_k\textrm{ , }[K_i,L_j]=i\e_{ijk}K_k\textrm{ , }[K_i,N_j]=i\d_{ij}K_4\textrm{ , }\nonumber\\
&[L_i,L_j]=i\e_{ijk}L_k\textrm{ , }[L_i,N_j]=i\e_{ijk}N_k\textrm{ , }\nonumber\\
&[N_i,N_j]=i\e_{ijk}L_k\textrm{ , }
\end{align}

so that if we define $2J^\pm_i\equiv L_i \pm N_i $, there are two commuting $SO(3)$ algebras\footnote{Of course this is a simple consequence of the isomorphism $SO(4)\sim SO(3)\times SO(3)$, and the fact that $SO(4)$ is a subgroup of the de Sitter group.}:
\begin{equation}
[J^a_i,J^b_j]=i\e_{ijk}J^a_k\delta^{ab}\nonumber\\
\end{equation}

It is plain to verify that the little group of a null vector is now the euclidean three-dimensional group $E(3)$, generated by the six elements:
\bea
&&[L_i,L_j]=i\e_{ijk}L_k\nonumber\\
&&[L_i,T_j]=i\e_{ijk}T_k
\eea
where $T_i\equiv K_i + N_i$, and the group that takes a null vector into a multiple of itself is none other than $SIM(3)$, the maximal subgroup of the de Sitter group (cf. \cite{Gibbons}), where the Lie algebra is augmented with the new generator $K_4$:
\begin{equation}
[K_4,T_i]=-iT_i
\end{equation}

It should be remarked that this is {\em not} equivalent to $\Lambda$VSR, which only included five generators, whereas we now enjoy seven, that is the cosmological constant just kills one generator with respect to the flat case. In the language appropiated for taking the flat limit $l\rightarrow \infty$, the $L_i$  are equivalent to our former $J_i$, so that we are keeping rotational invariance, $T_i\equiv K_i+P_i$ and $K_4$ is our former $H$. The theory breaks all boosts: $K_3$ is no longer a symmetry of the theory.\\

\section{VSR Fermion propagation in de Sitter}

Cohen and Glashow \cite{Cohen} proposed neutrino masses that neither violate lepton number nor require additional sterile states. The nonlocal Dirac equation would be
\be
\left(p\gsl-\frac{1}{2}m_{\nu}^2 \frac{n\gsl}{p\cdot n}\right)\nu_L=0\textrm{ ,}
\ee
where $n$ points to the prefered null direction. In this description all massive particles, including neutrinos, enjoy a standard dispersion relation $p^2=m^2$.\\

Lorentz violating effects are argued by the aforementioned authors to be of order $o(\g^{-2})$, except in a narrow cone about the privileged direction. Near the endpoint of the electron spectrum of beta decay, novel effects could be experimentally accesible.\\

The Cohen-Glashow equation can be formally derived from the local action for three chiral spinors:
\begin{equation}
\mathcal 
L=i\bar\nu\partial\gsl\nu+i\bar\chi\partial_n\psi+
i\bar\psi\partial_n\chi+im\bar\chi
n\gsl\nu+im\bar\psi\nu+\textrm{c.c.}
\label{accion}
\end{equation}

Although it seems at first sight that the symmetry of this lagrangian consists only on the euclidean group $E(2)$, that is, the set of transformations that leave invariant the null vector $n$, the symmetry is actually enhanced towards $SIM(2)$, provided fermions transform as:
\be
\chi_L\rightarrow \frac{1}{\l}D_s(\Lambda)\chi_L
\ee
where $D_s(\Lambda)$ is the usual $(0,1/2)$ spinor representation of the Lorentz group. The parameter $\l$ is the rescaling of $n$ caused by the $SIM(2)$ transformation. This is a new representation of $SIM(2)$, but not the one induced as a subgroup of the Lorentz group.\\

The VSR scheme can be seen partially (at least, for the $E(2)$ subgroup) as a dynamical consequence of this lagrangian, so coupling it to gravity will give us a new possibility to break the de Sitter invariance.\\

Let us now turn our attention to the description of the fermion propagation in de Sitter. We have reviewed in the Appendix for the convenience of the reader some facts on fermions in de Sitter space. Let us start from the lagrangian (written in a free falling local frame):
\begin{equation}\label{curvo}
\mathcal L=\sqrt{-g}\left(i\bar\nu\alpha^\mu\nabla_\mu\nu+i\bar\chi\nabla_n\psi+i\bar\psi\nabla_n\chi+im\bar\chi n\gsl\nu+im\bar\psi\nu+\textrm{c.c.}\right)\textrm{ ,}
\end{equation}
where $\alpha^\mu=\viel{}a a\mu\gamma^a$ is a set of curved gamma matrices, $n$ is a null vector field over the de Sitter space and $n\gsl$ stands for $g_{\mu\nu}\alpha^\mu n^\nu$. This seems the simplest generalization of the description we just introduced in flat space.\\

The Lagrangian defined using the Weyl technique (i.e., in a local inertial frame or tetrad) is diffeomorphism invariant, and enjoys local Lorentz invariance as well. It is natural to assume that the symmetry group of the action is the subgroup of the set of all isometries (de Sitter) that leave the null vector $n$ invariant:
\be
\pounds(k) n=0
\ee

In order to recover the flat space action, we would demand $n$ to be covariantly constant, $\nabla n=0$ which in fact implies that it is a Killing vector by itself, but there are not null Killing vectors over the de Sitter space.\\

Minimal coupling applied to $\partial n=0$ is then too restrictive. Let us assume instead:
\begin{equation}
\nabla n\to0\textrm{ if }l\to\infty
\end{equation}

In the riemannian coordinates, the covariant derivative of $n$ is:
\begin{equation}
\nabla_\mu n^\nu=\partial_\mu n^\nu+\frac{\Omega}{2l^2}(x_\mu n^\nu-x^\nu n_\mu+(n\cdot x)\delta^\nu_\mu)\textrm{ ,}
\end{equation}
so our flat space condition implies that $\partial_\mu n^\nu=0$ in this coordinates.\\

Let us now try to find the corresponding isometry subgroup. A general Killing field in de Sitter space has the following expression:
\begin{equation}
k^\mu=\lambda^\mu_{\phantom\mu\nu}x^\nu+a^\mu l\left(1+\frac{x^2}{4l^2}\right)-\frac{a\cdot x}{2l}x^\mu\textrm{ ,}
\end{equation}
where $\lambda$ is an infinitesimal (4 dimensional) Lorentz transformation and $a$ an infinitesimal 4 vector. If it commutes with $n$ then:
\begin{equation}
[n,k]^\mu=n^\nu\partial_\nu k^\mu=0=\lambda^\mu_{\phantom\mu\nu}n^\nu+a^\mu\frac{n\cdot x}{2l}-x^\mu\frac{n\cdot a}{2l}-n^\mu\frac{a\cdot x}{2l}
\end{equation}

This expression is a polynomial of degree 1 in $x$, so both terms must be zero:
\begin{equation}
a^\mu n_\nu-\delta^\mu_\nu(n\cdot a)-n^\mu a_\nu=0\to a\cdot n=0\textrm{ ,}
\end{equation}
so we should have $a\propto n$. In the other hand:
\begin{equation}
\lambda^\mu_{\phantom\mu\nu}n^\nu=0
\end{equation}
so $\lambda$ is a generator from $E(2)$.\\

This subgroup is precisely $E(2)\uplus\{P_+\}$, and it corresponds to one of our generalizations of the VSR group, namely, the first one, $\Lambda$VSR.

The equations of motion from the lagrangian (\ref{curvo}) read:
\begin{align}
&2\alpha^\mu\partial_\mu\nu+2\alpha^\mu\Gamma_\mu\nu-mn_\mu\alpha^\mu\chi-m\psi=0\nonumber\\
&2n^\mu\partial_\mu\psi+2n^\mu\Gamma_\mu\psi+mn_\mu\alpha^\mu\nu+\nabla_\mu n^\mu\psi=0\nonumber\\
&2n^\mu\partial_\mu\chi+2n^\mu\Gamma_\mu\chi+m\nu+\nabla_\mu n^\mu\chi=0
\end{align}

In stereographic coordinates, and assuming that the divergence of $n$ vanishes,
\begin{align}
&2\frac{\slashed\partial\nu}\Omega+\frac{3}{l^2} \slashed x\nu-m\Omega\slashed n\chi-m\psi=0\nonumber\\
&2\partial_n\psi-\frac{\Omega}{l^2}(\slashed x\slashed n-x\cdot n)\psi+m\Omega\nu=0\nonumber\\
&2\partial_n\chi-\frac{\Omega}{l^2}(\slashed x\slashed n-x\cdot n)\chi+m\nu=0
\end{align}

Let us now perform perform a formal  expansion of all variables in powers of 
the radius of the de Sitter space $l\equiv \frac{1}{M_\l}$, $\displaystyle\phi=
\sum_i\phi^i\frac1{l^i}$, including the vector $n$.\\

At zero order in $l$ we recover the flat space equations:

\begin{align}
&2\slashed\partial\nu^0-m\psi^0-m\slashed n^0\chi^0=0\nonumber\\
&2\partial_{n^0}\chi^0+m\nu^0=0\nonumber\\
&\partial_{n^0}\psi^0+m\slashed n^0\nu^0=0
\end{align}

At \emph{first} order in $l$, we have a non-trivial equations:
\begin{eqnarray}
&&2\slashed\partial\nu^1-m\psi^1-m\slashed n^0\chi^1-m\slashed n^1\chi^0=0\nonumber\\
&&\partial_{n^0}\chi^1+\partial_{n^1}\chi^0+m\nu^1=0\nonumber\\
&&\partial_{n^0}\psi^1+\partial_{n^1}\psi^0+m\slashed n^0\nu^1+m\slashed n^1\nu^0=0\nonumber
\end{eqnarray}
The fields that were auxiliary (i.e., with algebraic equations of motion) in the flat case
do propagate now, so that the correct procedure would be to use the Feynman rules to compute any
given process.
\par
Nevertheless, we can formally eliminate the auxiliary fields, just to get an idea of the 
differences  with respect to the flat case:
\begin{align}
&\psi^0=-\frac m2\frac{\slashed n^0}{\partial_{n^0}}\nu^0\nonumber\\
&\chi^0=-\frac m2\frac1{\partial_ {n^0}}\nu^0\nonumber\\
&\left(\slashed\partial+\frac{m^2}2\frac{\slashed n^0}{\partial_{n^0}}\right)\nu^0=0
\end{align}

The equations of motion to first order then read:
\begin{align}
&\psi^1=\frac1{\partial_{n^0}}\partial_{n^1}\frac1{\partial_{n^0}}\frac m2\slashed n^0\nu^0-\frac m2\frac1{\partial_{n^0}}\slashed n^1\nu^0-\frac m2\frac1{\partial_{n^0}}\slashed n^0\nu^1\nonumber\\
&\chi^1=\frac1{\partial_{n^0}}\partial_{n^1}\frac1{\partial_{n^0}}\frac m2\nu^0-\frac m2\frac1{\partial_{n^0}}\nu^1\nonumber\\
&\left(2\slashed\partial+\frac{m^2}2\{\slashed n^0,\frac1{\partial_{n^0}}\}\right)\nu^1+\frac{m^2}2\left(\{\slashed n^1,\partial_{n^0}\}-\{\frac1{\partial_{n^0}}\partial_{n^1}\frac1{\partial_{n^0}},\slashed n^0\}\right)\nu^0=0
\end{align}

The flat space limit ($l\rightarrow\infty$) is then plainly as above.\\

\section{Conclusions}
In the present paper we have generalized VSR to the physical situation in which a cosmological constant is present.\\

From the group theory approach, two different ways of breaking de Sitter invariance arise. The first, dubbed $\Lambda$VSR, in which we start from $SIM(2)$ and incorporate as many translations as possible (namely, only one). The remaining symmetry group is a five dimensional subgroup of the Poincar\'e group.\\

The second approach does {\em not} include $SIM(2)$ and is based upon $SIM(3)$, the subgroup that leaves a null vector proportional to itself. The symmetry group is in this case is seven-dimensional, that is only one generator short of the ordinary VSR when the cosmological constant vanishes.\\

Some lagrangians describing (free) fermion  propagation in each of the above-mensioned cases
were discussed, as well as the (dim) possiblity of detecting curvature effects in fermion
propagators. It seems that only in the case that, for some unknown reason,
the neutrino mass is of the order of the Hubble scale,
\be
m_\nu\sim M_\l
\ee
could those effects be (marginally) measurable.
The effects of the cosmological constant would be measured by experimentalists as 
Lorentz violating, position dependent,  mass terms. If there is in addition 
a breaking of de Sitter symmetry, there
are corrections near the endpoint of the beta decay spectrum proportional to the ratio
of the square of the mass, divided by the product of its momentum times its energy, 
$\frac{m^2}{E p}$
similar to the ones uncovered previously by Cohen and Glashow in the flat case.

\section*{Acknowledgments}
We are grateful to Andy Cohen and Patrick Meessen for illuminating remarks.
This work has been partially supported by the
European Commission (HPRN-CT-200-00148) and by FPA2003-04597 (DGI del MCyT, Spain), 
Proyecto HEPHACOS ; P-ESP-00346 (CAM) and  Consolider project  CSD 2007-00060 (MEC) 
({\em Physics of the Accelerated Universe}). R.V. is supported by a MEC grant, AP2006-01876.

\appendix
\section{Fermions in de Sitter}
There are two (equivalent) ways to consider fermion propagation in de Sitter space: the 
group theoretical one, first stated by Dirac \cite{Dirac} and the geometrical one
using free-falling frames, first proposed by Weyl \cite{Weyl}. They are well-known to 
be equivalent, at least at the level of the equations of motion\cite{Gursey}.\\

Let us consider physics in a local free-falling frame defined by a tetrad $\viel{}{a}{a}{\mu}$ \footnote{We refer to the exceedingly clear exposition in \cite{Ortin}}. Given a Dirac spinor with Lorentz transformation properties:
\begin{equation}
\psi'^\alpha=M_s(\Lambda)^\alpha_{\phantom\alpha\beta}\psi^\b\textrm{ ; }
\end{equation}
where the $(1/2,0)\oplus(0,1/2)$ (spinorial) representation of the Lorentz transformation $\Lambda$ is given by:
\begin{equation}
M_s(\Lambda)=e^{\frac14\sigma^{ab}\gamma_{ab}} \textrm{ , } 
\end{equation}
where $\g_{ab}\equiv \gamma_a\gamma_b-\gamma_b\gamma_a$. This field is assumed to be an scalar under diff-transformations:
\be
\d_D \psi=\xi^\a \pd_\a \psi
\ee

The covariant derivative of the spinor field is given by:
\begin{equation}
\nabla_\mu\psi=\left(\partial_\mu+\Gamma_\mu\right)\psi=\left(\partial_\mu-\frac14\omega_\mu^{\phantom\mu ab}\gamma_{ab}\right)\psi
\end{equation}
where the spin connection transforms as:
\begin{equation}
\delta_L\omega_\mu^{ab}=\partial_\mu\sigma^{ab}-\left[\omega_\mu,\sigma\right]^{ab}\textrm{ ; }
\end{equation}

A particular solution is given in terms of the Ricci rotation coefficients:
\begin{align}
&\Omega_{ab}^{\phantom{ab}c}=\viel {}aa\mu\viel{}bb\nu\partial_{[\mu}\viel c{\nu]}{}c\textrm{ ,}\nonumber\\
&\omega_{abc}=\Omega_{bca}-\Omega_{abc}-\Omega_{cab}\textrm{ ,}\nonumber\\
&\omega_{\m ab}\equiv e^c_{\phantom c\m} \omega_{cab}
\end{align}

This guarantees that under a  Lorentz transdormation $\s^{ab}(x)$,
\[
\d_L \nabla_\m \psi=\frac{1}{4}\s^{ab}\gamma_{ab}
\]
so that the term $\Phi(x)\equiv \bar{\psi}\g^\m \nabla_\m\psi$ transform as a worldsheet scalar, and the full action:
\be
S\equiv \int d^4 x \sqrt{|g|}\bar{\psi}\g^\m \nabla_\m\psi
\ee
is diff-invariant. This is true for a generic metric tensor.\\ 

When the spacetime enjoys isometries (as is the case for de Sitter space) then, for
those particular diffeomorphisms generated by the Killing vectors,
\be
\d_I g_{\m\n}=\pounds(k)g_{\m\n}=0
\ee
and the action is invariant under:
\be
\d_I \Phi=k^\a \nabla_\a \Phi
\ee
(owing to $\nabla_\a k^\a=0$).
The scalar $\Phi(x)$ can now be rewritten as:
\be
\Phi=\bar{\psi}\g^a e_a^\m\nabla_\m\psi
\ee
which conveys the invariance of the action under the transformations:
\be
\d_I \psi= k^\a \pd_\a\psi
\ee
{\em provided} the tetrad is chosed in such a way that $\d _I e_a^\m=0$. 

\par
When Riemann stereographic coordinates are used, i.e.:
\be
ds^2=\Omega^2 \eta_{\m\n}dx^\m dx^\n
\ee
with $\displaystyle\Omega=\frac{1}{1-\frac{x^2}{4 l^2}}$, $l=R_H$. The simplest choice for the tetrad one-form is plainly:
\be
e_a^{\phantom a\mu}=\Omega^{-1}\,\delta^\mu_a
\ee
The Dirac action with the former tetrad reads:
\be
S=\int d^4 x\,\Omega^4\,i\bar{\psi}\left(\frac{\slashed\partial}\Omega+\frac{3}{2 l^2}x\gsl\right)\psi
\ee
The effective-mass like term produced by the cosmological constant is very small, because it depends quadratically on the Hubble mass scale.


\end{document}